# Controlling Mn Depth Profiles in GaMnAs During High-Temperature Molecular Beam Epitaxial Growth

P. Xu, D. Qi, M.L. Ackerman, S.D. Barber, and P.M. Thibado[a]

Department of Physics, University of Arkansas, Fayetteville, Arkansas 72701

**Abstract**

Mn-doped GaAs thin films were grown at a high substrate temperature of 580 °C. During the growth process, the Mn cell temperature was ramped at different rates, resulting in a variety of different Mn concentration depth profile slopes, as measured using dynamic secondary ion mass spectrometry (SIMS). Results show that controlling the Mn deposition rate via temperature during molecular beam epitaxy (MBE) growth can mitigate the effect of Mn atoms diffusing toward the surface. Most importantly, the slope of the Mn concentration as a function of depth inside the sample can be tuned from negative to positive.

---

[a] Electronic mail: thibado@uark.edu

# 1. Introduction

In recent years, significant research efforts have been devoted to the potential use of dilute-magnetic semiconductors in spin-based electronics, or spintronics.[1] The reason for this is that, compared to conventional electronic devices, spin-based devices would have faster processing speeds and larger capacities for storing information.[2,3] Through these studies, it has been determined that some III-V magnetic semiconductors such as GaMnAs and InMnAs could be suitable for spintronics. For example, the discovery of ferromagnetism in Mn-doped GaAs grown by MBE[4] has led to interest in it as an ideal material for spintronics applications.[5,6]

The future success of Mn-doped GaAs in spintronics is dependent upon controlling where the Mn atoms position themselves within the GaAs host matrix. In principle, Mn can be incorporated into substitutional Ga or As positions,[7,8] and Mn atoms can occupy interstitial sites surrounded by four Ga or As atoms.[9] When Mn substitutes Ga atoms, the Mn ions provide local magnetic moments and holes that mediate long-range ferromagnetic coupling between the local moments.[10] Interstitial Mn atoms, on the other hand, can adversely affect the Curie temperature of GaMnAs.[11,12] They act as double donors and compensate the substitutional Mn acceptors to mediate the ferromagnetic interaction between Mn spins.[13]

In addition, another key factor in the success of Mn-doped GaAs for use in spintronics is the distribution of the Mn ions throughout the film layers. Using SIMS, Xu *et al.* showed that growth at a substrate temperature of 580°C for GaMnAs results in the Mn diffusing to the surface of the film during growth, leading to a non-uniform Mn concentration profile.[14] The effect is not small; the concentration in some instances drops by a factor of 1000 from the surface of the thin film to the internal bottom edge of the thin film (which is only one micron deep).

To eliminate this non-uniformity, the majority of research in the GaMnAs system is now done using low-substrate temperature growth conditions. However, low-temperature growth poses its own problems. These were quantified by J. Mašek *et al.* using high-resolution X-ray. They determined a single formula which models the concentration of Mn on Ga sites, As on Ga sites, and Mn on interstitial sites.[15] The concentration of As on Ga sites is known to result from low-temperature growth. To study its role further, J. Sadowski and J. Domagala tuned the As to Ga ratio during growth to minimize the concentration of As antisite defects and, in turn, improved the crystalline quality.[16] Furthermore, S. Mack *et al.* correlated the reduction in the concentration of As antisite defects with the improvement in the concentration of Mn on Ga sites, as well as an increase in the critical temperature.[17] The origin of As antisite defects has been well documented; they can be eliminated by growing GaAs at high substrate temperatures in combination with an $As_4$/Ga flux ratio between 10 and 20.[18] To offer new strategies for dealing with these problems, in this article we describe our study which revisited growing GaMnAs at 580 °C with the goal of controlling the Mn concentration profile.

## 2. Experimental details

Samples were prepared in a Riber 32 ultrahigh vacuum (UHV) (~2 × $10^{-10}$ Torr) MBE growth chamber which combines Ga and Mn effusion cells with a two-zone As valved-cracker cell and a reflection high-energy electron diffraction (RHEED) system operating at 1.5 keV. Commercially available, "epiready," semi-insulating, 2-inch-diameter GaAs(001) ±0.1° wafers were cleaved into quarters, then indium-mounted on a 2-inch-diameter standard MBE molybdenum block. The substrate was heated to 580 °C, concurrently exposing the surface to 10 μTorr $As_4$ to remove the surface oxide layer. Before growth the Ga and terameric As beam equivalent pressures (BEP) were set and measured. The As$_4$ BEP was set to 1.0 µTorr and the Ga BEP was set to give an As$_4$/Ga flux ratio of ≈10. Then a thin buffer layer of GaAs was grown on the substrate for 5 min. During this process, RHEED oscillations determined that the growth rate of the GaAs was ~650 nm/h. Finally, the substrate temperature was maintained at 580 °C and GaMnAs films were grown for ~1.7 h.

During the growth process, the key differentiation made between samples was the method by which the Mn cell temperature was controlled. To illustrate, the cell temperature for each of the three GaMnAs films grown is plotted as a function of growth time in Fig. 1. For reference, the corresponding post-growth sample thickness in nanometers is shown at the top of the plot, with the zero nm position representing the surface of the final thin film. The cell temperature was controlled during the growth using a proportional-integral-derivative (PID) controller (Eurotherm), which has a precision of ±0.1 °C. For the first sample, plotted with daggers, the cell temperature was held constant at 900 °C throughout the growth process. The ramping rate, $r$, for the Mn cell is labeled as 0.0 °C/min for this growth condition. For the second sample, plotted with squares, the Mn cell temperature began at 900 °C and was decreased at a constant ramping rate of $r$ = 1.0 °C/min throughout the growth period to a final temperature $T_f$ = 800 °C. For the third sample, plotted with triangles, the cell temperature also began at 900 °C, but was decreased at a constant rate of $r$ = 1.5 °C/min throughout the growth period to a final temperature $T_f$ = 750 °C. For the remainder of this paper, the samples will be differentiated based on the Mn cell temperature ramping rate, $r$, or the Mn cell temperature reached at the end of their respective growth period, $T_f$.

Upon completion of the sample growth, the substrate was cooled at a rate of 1.5 °C/s to 200 °C, removed from the UHV system, then cleaved into multiple smaller pieces (10 × 10 mm$^2$) for characterization measurements. The Mn concentration profiles were then determined using a dynamic SIMS (Perkin-Elmer PHI 6300 Quadrupole). This process involved sputtering the samples with an argon ion gun, using a focused spot-size of about 1 mm in diameter. The depths of the sputtered craters were measured using a profilometer (Tencor Alphastep 200). It was assumed that the average sputtering rate of $Ga_xMn_{1-x}As$ did not change with the Mn concentration. Finally, Auger experiments were performed on the samples to confirm the calibration of the Mn concentration as determined by SIMS.

## 3. Results

The SIMS Mn concentration depth profiles for the three GaMnAs thin films are shown in Fig. 2. The top curve, plotted with daggers, shows a film grown with a constant 900 °C Mn cell temperature. This film exhibited the greatest variation across the sample depth, with a Mn concentration of $10^{23}$ cm$^{-3}$ at the surface, dropping exponentially to a concentration of $10^{21}$ cm$^{-3}$ at a depth of 1.1 μm. The second film, plotted with squares, was prepared with a Mn cell temperature ramping down at $r = 1.0$ °C/min. This film exhibits the most uniform Mn concentration profile, decreasing from $3\times10^{21}$ cm$^{-3}$ near the surface to $4\times10^{20}$ cm$^{-3}$ at a depth of 1.1 μm. The final film, plotted with triangles, was prepared with $r = 1.5$ °C/min. Unlike the other two samples, this film displayed a gradual increase in Mn concentration with depth, about $2\times10^{19}$ cm$^{-3}$ near the surface and rising to approximately $4\times10^{20}$ cm$^{-3}$ at a depth of 1.1 μm.

To quantify the Mn deposition rate versus Mn cell temperature, additional Mn-doped GaAs thin films were grown using the method discussed earlier, but with a substrate temperature of 250 °C. The Mn cell temperature was held constant at 900, 800, and 750 °C during the growth of these calibration samples, with SIMS analysis providing the Mn concentration. This analysis showed a uniform Mn concentration throughout most of the GaMnAs thin film. The Mn deposition rate (shown as concentration) as a function of Mn cell temperature is displayed in Fig. 3, showing that, as expected, the rate increases exponentially with cell temperature. It should be noted that the pre-factor is $10^{30}$ cm$^{-3}$, while the activation energy is 1.5 eV.

## 4. Discussion

A primary outcome of this study was the demonstration of the tunability of the slope of the Mn concentration, as illustrated in Fig. 2. Most exciting was the discovery that the Mn concentration profile can be made nearly flattened by systematically reducing the Mn cell temperature during the growth of the thin film. This demonstrates that the Mn which floats up from below can be re-incorporated into the growing film, providing evidence that the Mn that floats up is no different, chemically-speaking, than freshly deposited Mn emanating from the Knudsen cell. This insight offers clarification and simplification as one cannot distinguish between the freshly deposited Mn and the Mn which has floated up.

A linearly-decreasing Mn cell temperature during the growth period appears to reduce the effect of the Mn as it floats up. A linearly decrease in time results in an exponentially-decaying Mn deposition rate, as was illustrated in Fig. 3. Correspondingly, the Mn concentration grows exponentially when there is a constant deposition rate in time (see dagger plot in Fig. 2). Thus, it can be surmised that the two effects, namely Mn atoms floating up and a decreasing Mn deposition rate, cancel each other out and create a uniform Mn concentration profile. It is not yet clear what ramping rate is appropriate for the temperature of the Mn cell in every situation. In general, however, a constant ramping rate of $r = 1.0$ °C/min throughout the growth period creates the best profile. It is worth noting that this ramping rate decreases the deposition of the Mn by about a factor of ten from the beginning of the growth to the end.

From previous high substrate-temperature studies, it has been shown that the slope of the Mn concentration profile due to Mn floating along the growth front is nearly independent of the Mn cell temperature.[14] This suggests that the "floating up" rate is not dependant upon the deposition rate, but that it is simply a random chance event for each Mn atom. Further, this implies that the Mn floats up at a fixed rate, as shown in the lowest curve in Fig. 2. Here, the ramping rate of $r = 1.5$ °C/min. throughout the growth process shows that the Mn incorporation still occurs; however, the concentration actually drops as it approaches the surface of the film. Therefore, with the appropriate choice of Mn cell starting temperature, it is possible to control the final Mn concentration profile of the GaAs thin film. To illustrate, starting the Mn cell at 850 °C or 950 °C, but continuing to ramp down at a rate of $r = 1.0$ °C/min, a nearly uniform Mn concentration profile would result (with a lower or higher overall concentration, respectively).

## 5. Conclusions

In summary, the goal of this study was to grow Mn-doped GaAs 580 °C and then, within this constraint, determine a growth algorithm for controlling the depth profile for the Mn concentration. We were successful at finding the growth algorithm. Namely, by lowering the Mn cell temperature linearly in time, the Mn flux decreased exponentially during the growth of the thin film. This decreasing flux was precisely tuned to cancel the exponentially increasing Mn concentration which normally occurs due to it floating up along the growth front. In particular, a decreasing ramping rate of ~1 ºC/min. for a growth rate of ~1 monolayer/second appears to be optimal. Ultimately, because of the mechanism behind the Mn floating up process, it has been shown that a proper setting of the Mn depositing ramp rate can yield a uniform distribution of Mn throughout GaAsMn film. In fact, this method creates the ability to control the slope of the Mn concentration profile and tune it from negative to positive.

Many algorithms remain to be investigated, for example growing this material at different growth rates, at different Mn concentrations, with surfactants, etc. Future studies will also need to determine the concentration of Mn on Ga sites versus interstitial sites. This study does not address this, and only focuses on determining a growth algorithm for controlling the doping depth profile. Also, note that the growth algorithm developed in this study does not offer a solution for short period superlattices, where the major concern is the Mn diffusing away from the region it was deposited on and toward the regions where it was not deposited.[19] Nevertheless, in this study the problem of the exponential increase in the Mn concentration during high-substrate temperature growth has been studied, understood, and addressed.

## Acknowledgment

This work was supported in part by the National Science Foundation (NSF) under grant number DMR-0855358 and the Office of Naval Research (ONR) under grant number N00014-10-1-0181.

-10-

[14] J. F. Xu, P. M. Thibado, C. Awo-Affouda, F. Ramos, and V. P. Labella, J. Vac. Sci. Technol. B **25**, 1476 (2007).

[15] J. Mašek, J. Kudrnovský, and F. Máca, Phys. Rev. B **67**, 153203 (2003).

[16] J. Sadowski, and J. Z. Domagala, Phys. Rev. B **69**, 075206 (2004).

[17] S. Mack, R. C. Myers, J. T. Heron, A. C. Gossard, and D. D. Awschalom, Appl. Phys. Lett. **92**, 192502 (2008).

[18] X. Liu, A. Prasas, J. Nishio, and E. R. Weber, Appl. Phys. Lett. **67**, 279 (1995).

[19] M. Poggio, R. C. Myers, N. P. Stern, A. C. Gossard, and D. D. Awschalom, Phys. Rev. B **72**, 235313 (2005).

**Figure Captions**

Fig. 1. Mn cell temperature ramping sequences for the three Mn-doped GaAs thin films are shown. The corresponding post-growth sample depth is shown on the top of the graph for easy reference. The cell temperature during the growth of the first sample is shown in the top curve and is plotted using daggers. For this sample, the Mn cell temperature was held constant at 900 °C throughout, and is therefore labeled with a ramping rate $r = 0.0$ °C/min and final cell temperature $T_f = 900$ °C. For the second sample, shown with squares, the Mn cell temperature was ramped down at a rate $r = 0.0$ °C/min giving a final cell temperature $T_f = 800$ °C. The third sample, shown with triangles, has a Mn cell temperature ramp rate $r = 1.5$ °C/min giving a final cell temperature $T_f = 750$ °C.

Fig. 2. The Mn concentration depth profiles determined with SIMS for the three GaMnAs samples are shown above. The profile plotted with daggers was the sample grown with a constant Mn cell temperature of 900 °C throughout and is labeled as $r = 0.0$ °C/min. The profile plotted with squares was grown with the Mn cell temperature ramping at $r = 1.0$ °C/min to the final cell temperature $T_f = 800$ °C. The profile plotted with triangles was grown with the Mn cell temperature ramping at 1.5 °C/min, giving a final cell temperature $T_f = 750$ °C. Note that the $T_f = 900$ °C sample is slightly thicker than the other two samples due to substantially extra amount of Mn in the film.

Fig. 3. Three Mn concentrations for various GaMnAs thin films grown using a low substrate temperature of 250 °C, and different but constant Mn cell temperatures (equal to 900 ºC, 800 ºC and 750 ºC) are shown above. The Mn concentrations were measured using SIMS. The fitting curve illustrates the exponential dependence of the Mn concentration on the Mn cell temperature.

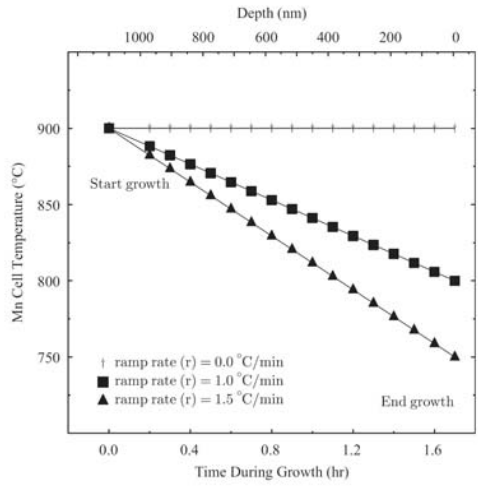

Figure 1.

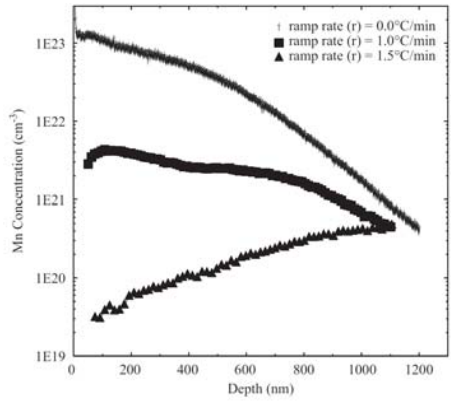

Figure 2.

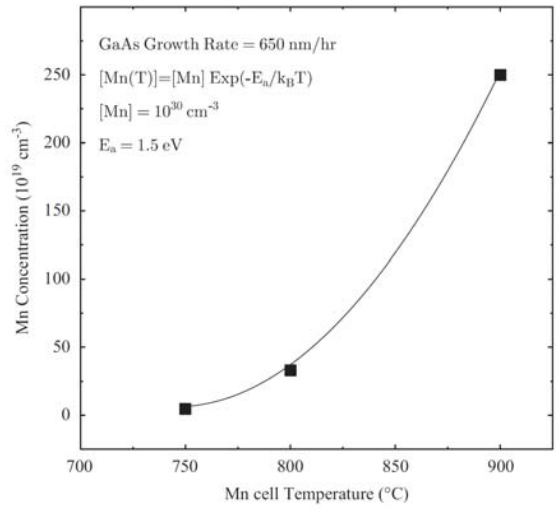

Figure 3.